\documentclass[twocolumn,preprintnumbers,amsmath,amssymb]{revtex4}
\usepackage{graphicx}
\usepackage{dcolumn}
\usepackage{bm}

\begin{document}

\title{The breakdown of the zeroth law of thermodynamics and the definition of temperature in small quantum systems}
\author{Pei Wang}
\email{wangpei@zjut.edu.cn}
\affiliation{Institute of applied physics, Zhejiang University of Technology, Hangzhou, P. R. China
}

\begin{abstract}
We study two small quantum systems coupled to the same reservoir which is in thermal equilibrium. By studying the particle density and the energy density in the two systems before and after they contact each other, we find that the two systems are not in thermal equilibrium with each other. Our result shows that the zeroth law of thermodynamics is broken in small quantum systems at low temperatures. Therefore, the traditional way of defining temperature fails due to the breakdown of the transitive relation of thermal equilibrium. Then we show a different way of defining temperature by attaching an auxiliary site, which plays the role of a thermometer, to the small quantum system.
\end{abstract}

\maketitle

\section{introduction}

Thermodynamics, developed in the 19th century, contains four fundamental laws. Among them the zeroth law of thermodynamics states the transitive relation of thermal equilibrium and is the prerequisite of the existence of temperature. For more than one hundred years, it is believed to be a fundamental law of nature and valid throughout various systems, either classical or quantum. 

The zeroth law of thermodynamics states that two systems in thermal equilibrium with a third one are also in equilibrium with each other. In other words, the thermal equilibrium is a transitive relation, so that one can define a physical quantity called temperature for a set of systems which are in equilibrium with each other. This statement can be checked theoretically by studying the dynamics of three systems coupled together. In detail, we suppose two systems coupled to a large thermal reservoir at the temperature $T$. At the initial time, the two systems are isolated to each other and are both in equilibrium with the reservoir. According to the zeroth law of thermodynamics, they must also be in equilibrium with each other. This can be checked by switching on a coupling between the two systems. If they have been in equilibrium, there should be no exchange of energy or particles between them after the coupling is switched on. Conversely, if switching on the coupling causes the exchange of energy or particles, which can be shown by the changing of energy density or particle density in the two systems, the zeroth law breaks down. 

In this paper, we check the zeroth law in small quantum systems which obey the quantum laws. In recent years, with the development of the nanotechnology intense efforts have been devoted to study the thermalization process in different small quantum systems. Studies show that an open small quantum system may not evolve into an equilibrium state with a canonical density matrix as the macroscopic systems~\cite{goldstein09,wangwenge,hartmann04,znidaric,cazalilla,rigol06,rigol07,rigol08,kollath,manmana,barthel08,cramer,eckstein,gelin,hershfield93,kollar,wang}. This result challenges the method of measuring the temperature. Because the standard way of measuring the temperature of a system is to attach a thermometer to it. The thermometer will then evolve into an equilibrium state with the temperature as same as the system, which can then be readout. However, if the thermometer does not thermalize, the readout of it will depend upon how it is coupled to the system and then lose the meaning of temperature.

In this paper we will show that not only the method of measuring temperature but also the definition of temperature is challenged in small quantum systems. The traditional definition of temperature is based on the zeroth law of thermodynamics, which will be shown not to stand up in small quantum systems. We will introduce a different way of defining the temperature, which circumvent the zeroth law by attaching an auxiliary system to the small quantum system. 

The paper is organized as follows. In Sec.~II, we introduce the main model in this paper, an electron reservoir coupled to two small systems, and the method of solving it. In Sec.~III, we discuss the breakdown of the zeroth law. A method of defining the temperature in small quantum systems is introduced in Sec.~IV. In Sev.~V, we extend our discussion in a model where the number of particles is conserved. Sec.~VI is a short summary.

\section{Model and method}

We employ two chains consisting of $n$ sites respectively to model the two small quantum systems, which are coupled to an electron reservoir at the temperature $T$. At initial time the two chains are isolated to each other. To see if they have been in thermal equilibrium with each other, we switch on a coupling between the end sites of the two chains. The Hamiltonian describing the chains and the reservoir is expressed as
\begin{eqnarray}\nonumber\label{chainhamiltonian}
\hat H &=& \sum_{k\sigma} \epsilon_k \hat c^\dag_{k\sigma} \hat c_{k\sigma} +  \sum_{\alpha k \sigma} ( h_\alpha \hat c^\dag_{k\sigma} \hat c_{\alpha 1 \sigma} + h.c.) \\ && \nonumber + \epsilon \sum_{\alpha\sigma,j=1}^n \hat c^\dag_{\alpha j \sigma} \hat c_{\alpha j \sigma} +  \sum_{\alpha \sigma,j=1}^{n-1} g_\alpha( \hat c^\dag_{\alpha j\sigma} \hat c_{\alpha, j+1, \sigma} + h.c.) \\ && + d ( \hat c^\dag_{L n \sigma} \hat c_{R n \sigma} + h.c.),
\end{eqnarray}
where $\alpha=L,R$ denotes the two chains, $\sigma=\uparrow,\downarrow$ the spin, $\hat c_{k\sigma}$ the annihilation operator of the electron in the reservoir, $\hat c_{\alpha j \sigma}$ the annihilation operator at site $j$ in the chain $\alpha$, $h_\alpha$ the coupling between the chain $\alpha$ and the reservoir, $g_\alpha$ the coupling between two neighbor sites in the chain $\alpha$ and $d$ the coupling between two chains.

The initial correlation is absent in this model. So the Keldysh Green's function approach~\cite{datta,jauho} can be employed for calculating the electron density in the two chains before and after the coupling $d$ is switched on.

The spin-dependent electron density at arbitrary site $m$ in the chain $\alpha$ can be expressed by the lesser Green's function as
\begin{eqnarray}\label{inversefourier}
n_{\alpha m} = \frac{1}{2\pi i} \int_{-\infty}^\infty d\omega  G^<_{\alpha m,\alpha m}(\omega).
\end{eqnarray}
The $G^<_{\alpha m,\alpha m}$ is the element of a $2n\times 2n$ Green's function matrix, which according to the Keldysh formalism can be related to the retarded and advanced Green's function matrix by
\begin{eqnarray}\label{lesserkeldysh}
 G^<= G^r \Sigma^< G^a.
\end{eqnarray}
In this formula, the lesser self-energy matrix contains only four non-zero elements, which are listed below:
\begin{eqnarray}\label{lesserselfenergy}
\big\{ \begin{array}{c} \Sigma^<_{L 1,L 1}=2i\Gamma_L f(\omega) \\ \Sigma^<_{R 1,R 1}=2i\Gamma_R f(\omega) \\ \Sigma^<_{L 1,R 1}=\Sigma^<_{R1,L1}=2i\Gamma_{LR} f(\omega) \end{array} .
\end{eqnarray}
Here $f(\omega)= \displaystyle \frac{1}{e^{\frac{\omega}{k_B T}} + 1}$ denotes the Fermi distribution function, $\Gamma_\alpha = \rho \pi h_\alpha^2$ and $\Gamma_{LR} = \rho \pi h_L h_R$ denote the level broadening at the edge sites due to the coupling to the reservoir and $\rho$ is the density of states in the reservoir. Substituting Eq.~\ref{lesserselfenergy} into Eq.~\ref{lesserkeldysh}, we obtain 
\begin{equation}
\begin{split}
 G^<_{\alpha m,\alpha m} =& 2i f \left(|G^r_{\alpha m,L1}|^2\Gamma_L+|G^r_{\alpha m,R1}|^2\Gamma_R \right. \\ & \left.
+ 2\Gamma_{LR}\textbf{Re} (G^r_{\alpha m,L1}G^{r*}_{\alpha m,R1} ) \right),
\end{split}
\end{equation}
where we use the relation $G^{a} = G^{r\dag}$. Then the electron density can be written as
\begin{equation}\label{expressiondensity}
 \begin{split}
 n_{\alpha m} =  \frac{1}{\pi } \int_{-\infty}^\infty d\omega  f & \left(|G^r_{\alpha m,L1}|^2\Gamma_L+|G^r_{\alpha m,R1}|^2\Gamma_R \right. \\ & \left.
+ 2\Gamma_{LR}\textbf{Re} (G^r_{\alpha m,L1}G^{r*}_{\alpha m,R1} ) \right).
\end{split}
\end{equation}

All the $2n\times 2n$ retarded Green's functions are the elements of a matrix $G^r$, which satisfies the Dyson's equation:
\begin{eqnarray}\label{dyson}
 G^r(\omega)= G^{0r}(\omega)+ G^{0r}(\omega) \Sigma^r(\omega) G^{r}(\omega).
\end{eqnarray}
Here $G^{0r}(\omega)$ is the free Green's function matrix as $h_\alpha=g_\alpha=d=0$, and its elements are 
\begin{eqnarray}\label{g0rdef}
{G}^{0r}_{\alpha j,\alpha' j'}(\omega) = \delta_{j,j'} \delta_{\alpha,\alpha'} \frac{1}{\omega- \epsilon+i\eta},
\end{eqnarray}
where $\eta$ is an infinitesimal number. And $\Sigma^r(\omega)$ is the self-energy matrix, whose elements come from the kinetic energy of electrons hopping between neighbor sites and the level broadening at the edge sites. The subdiagonal and superdiagonal elements of $\Sigma^r(\omega)$ are
\begin{eqnarray}\label{tridiagonalelement}
\Sigma^r_{\alpha j,\alpha j+1} = \Sigma^r_{\alpha j+1,\alpha j} = g_\alpha. 
\end{eqnarray}
The coupling between the left and right chain gives
\begin{eqnarray}
\Sigma^r_{L n,R n} = \Sigma^r_{R n,L n} = d. 
\end{eqnarray}
And the level broadening at the edge sites gives
\begin{eqnarray}\label{edgeelement}
\big\{ \begin{array}{c} \Sigma^r_{\alpha 1,\alpha 1}(\omega) = -i\Gamma_\alpha \\
 \Sigma^r_{L 1,R 1}(\omega) = \Sigma^r_{R 1,L 1}(\omega)= -i\Gamma_{LR} \end{array} . 
\end{eqnarray}
Solving the Dyson's equation, we obtain the retarded Green's functions:
\begin{eqnarray}\label{dysonsolution}
 G^r = (G^{0r-1}-\Sigma^r)^{-1}.
\end{eqnarray}
Finally, the electron density is got by substituting Eq.~\ref{dysonsolution} into Eq.~\ref{expressiondensity}.

In the above formalism, the electron-electron interaction is not considered. The interaction can be included by adding a Hubbard term $\hat H_U = U \sum_{\alpha j} \hat c^\dag_{\alpha j \uparrow} \hat c_{\alpha j\uparrow}  \hat c^\dag_{\alpha j \downarrow} \hat c_{\alpha j\downarrow} $ to the Hamiltonian. In the self-consistent mean field approximation, the interacting model can be easily solved by replacing the on-site potential $\epsilon$ by the $\epsilon_{\alpha j} = \epsilon+ U n_{\alpha j}$ in the above formalism and then calculating the electron density $n_{\alpha j}$ self-consistently. This has been proved to be a good approximation as the interaction $U$ is small.

\section{The breakdown of the zeroth law of thermodynamics in small quantum systems}

First we consider the special case that each chain contains only a single site and the interaction $U=0$. Now the chain $L$ and $R$ are also called the site $L$ and $R$ respectively. The spin-dependent electron density at the site $L$ is
\begin{eqnarray}
 n_{L} = \frac{1}{\pi} \int d \omega \frac{\left( \sqrt{\Gamma_L} (\omega-\epsilon) + d \sqrt{ \Gamma_R} \right)^2 }{|A|^2} f(\omega),
\end{eqnarray}
and that at the site $R$ is
\begin{eqnarray}
 n_{R} = \frac{1}{\pi} \int d \omega \frac{ \left( \sqrt{\Gamma_R} (\omega-\epsilon) + d \sqrt{ \Gamma_L}\right)^2}{|A|^2} f(\omega),
\end{eqnarray}
where $A=( \omega-\epsilon +i\Gamma_L)(\omega-\epsilon +i\Gamma_R)- (i\Gamma_{LR}-d)^2$. 

\begin{figure}
\includegraphics[width=0.45\textwidth]{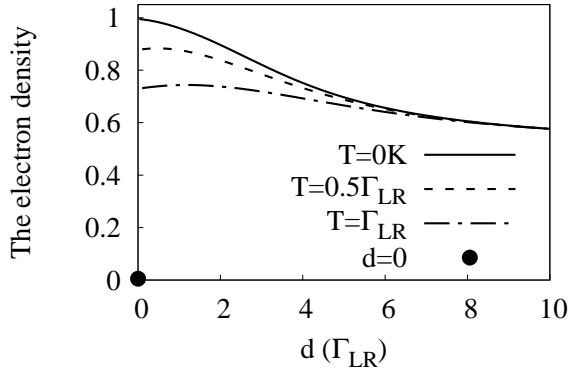}
\caption{The electron density at the weakly-coupled site $R$ as a function of the coupling $d$ at different temperatures. The Fermi energy is set to be the energy zero. The level broadening at the two sites is $\Gamma_L=10\Gamma_{LR}$ and $\Gamma_R=0.1\Gamma_{LR}$ respectively, and the energy level is $\epsilon=-\Gamma_{LR}$. The black circle denotes the electron density when the two sites are isolated to each other ($d=0$). }
\end{figure}
The electron density at the sites depends strongly upon the coupling $d$ as the level broadening of the two sites are very different. We set the Fermi energy of the reservoir to be the energy zero. In the case of the on-site energy level $\epsilon<0$ and $\Gamma_L\gg |\epsilon| \gg \Gamma_R$, the electron density at the strongly-coupled site (the site $L$) is $n_L=0.5$ before the two sites contact each other. And the electron density at the weakly-coupled site (the site $R$) is $n_R=0$, since it is screened. Switching on a coupling between the two sites has little effect on the electron density at the strongly-coupled site. But the electron density at the weakly-coupled one will change abruptly into a finite value depending upon the temperature. As $d$ increasing the electron density at both sites will go towards $0.5$ (see Fig.~1). 

We see that the electrons at the two sites will redistribute themselves after the sites contact each other. So the two sites which are in equilibrium with a reservoir are not in thermal equilibrium with each other. The transitive relation of thermal equilibrium is broken in this case.
\begin{figure}
\includegraphics[width=0.45\textwidth]{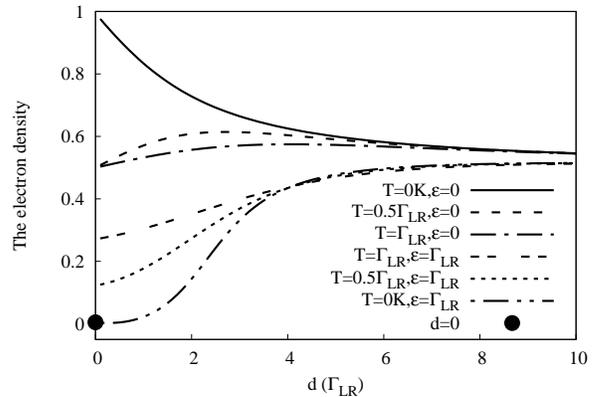}
\caption{The electron density at the weakly-coupled site $R$ as a function of the coupling $d$ at different temperatures and the energy levels $\epsilon$. The level broadening is $\Gamma_L=10\Gamma_{LR}$ and $\Gamma_R=0.1\Gamma_{LR}$. The energy level $\epsilon$ takes two possible values: $0$ and $\Gamma_{LR}$. And the temperature is chosen as $T=0K,0.5\Gamma_{LR}$ and $\Gamma_{LR}$. The black circle denotes the electron density when the two sites are isolated to each other ($d=0$). }
\end{figure}
This picture will not change as $\epsilon \geq 0$ (see Fig.~2). In a conclusion, the electron density at the weakly-coupled site will change significantly after the two sites contact each other whatever the temperature and the on-site energy level are. The two sites are not in equilibrium as their level boadening is much different from each other. 

The breakdown of the zeroth law is because the quantum correlation between the two sites is sensitive to the coupling strength $d$. But as the chain length increasing, the coupling at the boundary will become unimportant. And then the zeroth law should recover. This is verified by the numerical calculations. 

\begin{figure}
\includegraphics[width=0.45\textwidth]{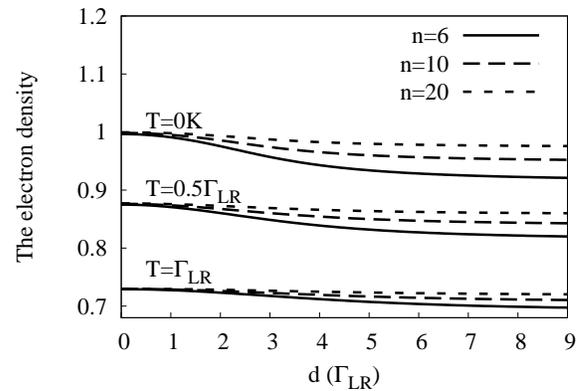}
\caption{The averaged on-site electron density in the weakly-coupled chain as a function of $d$ at different temperatures. The temperature is chosen as $T=0K, 0.5\Gamma_{LR}$ and $\Gamma_{LR}$. The length of the chains varies from $6$ to $20$. The chain $L$ is strongly-coupled to the reservoir with $\Gamma_L=g_L=10\Gamma_{LR}$, while the chain $R$ is weakly-coupled to the reservoir with $\Gamma_R=g_R=0.1\Gamma_{LR}$. The energy level is set to be $\epsilon=-\Gamma_{LR}$. As the chain length increasing, the dependence of the electron density on $d$ is reduced.}
\end{figure}
We study two chains which are coupled to the reservoir in the very different strength, i.e., $\Gamma_L \gg \Gamma_R$. The averaged electron density in the weakly-coupled chain as a function of $d$ is shown in Fig.~3 for different chain length. We find that the electron density changes with the coupling $d$. But the change becomes small as the chain length $n$ increasing. 

And at the finite temperature, the thermal fluctuation will suppress the influence of the coupling between the two chains. This is clear in the case of both $n=1$ and $n>1$ (see Fig.~1, 2 and 3). The breakdown of the zeroth law is only significant at low temperatures.

\begin{figure}
\includegraphics[width=0.45\textwidth]{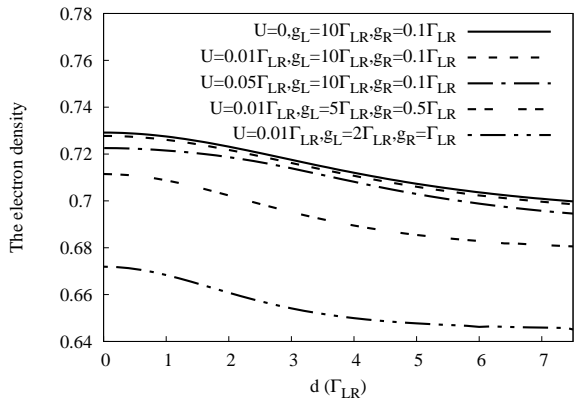}
\caption{The averaged on-site electron density in the weakly-coupled chain as a function of $d$ at different $U$, $g_L$ and $g_R$. The temperature is chosen as $T=\Gamma_{LR}$. The length of the chains is $n=6$. And the coupling between the chains and the reservoir is $\Gamma_L=10\Gamma_{LR}$ and $\Gamma_R=0.1\Gamma_{LR}$ respectively. The energy level is set to be $\epsilon=-\Gamma_{LR}$. }
\end{figure}
We study the effect of the interaction $U$. As shown in Fig.~4, the interaction changes the electron density. However, what we are interested in is whether switching on a coupling $d$ will change the electron density. This is found to be independent to the interaction. And we also find that the hopping $g_L$ and $g_R$ in the two chains is not important in our picture. 

In a word, the transitive relation of thermal equilibrium breaks down in short chains at low temperatures. This indicates that the traditional method of defining temperature fails for small quantum systems at low temperatures. It is necessary to find a new definition of temperature in this situation as the correlation at the boundary cannot be neglected. As will be shown next, such a definition can be obtained by attaching an auxiliary site to the system that we want to measure. We also find that the zeroth law is recovered when we increase either the temperature or the system size. So the zeroth law is always a good approximation in everyday's life.

\section{The local temperature in a small quantum system}

Due to the breakdown of the zeroth law in small quantum systems at low temperatures, we introduce an alternate way of defining the temperature, which is inspired by the process of using a mercury thermometer to measure the temperature of an object. The measuring process is to attach the thermometer to the target and wait for a long time, so that an equilibrium between the two systems is established. The temperature of the target is as same as that of the thermometer, which can then be obtained from the relation between the temperature and the volume of the mercury. 

The condition that this procedure works is that the thermometer after attached to the target should be in a canonical state. This can be realized when the coupling between the thermometer and the target is infinitesimal. An example showing this is in the Anderson impurity model without interaction, which describes an impurity site coupled to a Fermi sea at the temperature $T$ and the chemical potential $\mu$. The impurity site plays the role of the thermometer. As is well known, the electron density at the impurity site goes to the Fermi function
\begin{eqnarray}\label{occupation}
n_d =\displaystyle \frac{1}{e^{\frac{1}{k_B T}(\epsilon_d -\mu)}+1},
\end{eqnarray}
in the limit as the coupling between the site and the Fermi sea goes to zero. Here $\epsilon_d$ denotes the energy level of the impurity site. This fact indicates a method of defining the temperature of a small quantum system at low temperatures.

We define the local temperature of a quantum system in thermal equilibrium by attaching an auxiliary site to the system. The auxiliary site plays the role of the thermometer. The coupling between the auxiliary site and the quantum system is switched on at the time $t=0$. As the coupling goes to zero, the electron density at the auxiliary site as a function of its energy level must be the Fermi function when the time goes to infinity. According to Eq.~\ref{occupation}, the electron density $n_d(\epsilon_d)$ satisfies
\begin{eqnarray}\label{linearrelation}
 \ln \left(\frac{1}{n_d}-1 \right)= \frac{1}{k_B T} \epsilon_d -\frac{ \mu}{k_B T} .
\end{eqnarray}
Then the temperature $T$ and the chemical potential $\mu$ of the quantum system can be determined by plotting $ \ln (1/n_d-1)$ with respect to $\epsilon_d$. The slope of this linear function is just the inverse of the temperature $\frac{1}{k_B T}$. 

The temperature that we obtain in this way is the local temperature, which depends upon the position where the auxiliary site is coupled. For a system in thermal equilibrium, the local temperature is the same everywhere, equaling to the global temperature. 

\begin{figure}
\includegraphics[width=0.45\textwidth]{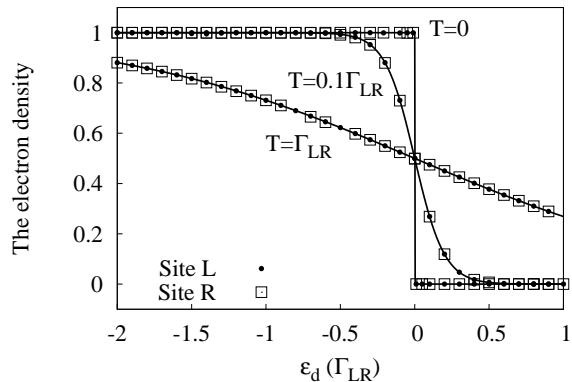}
\caption{The spin-dependent electron density $n_d$ at the auxiliary site as a function of its energy level. The black circle and the white square denote the electron density as the auxiliary site is coupled to the site $L$ and $R$ respectively. The site $L$ and $R$ is coupled to a Fermi sea in equilibrium with the level broadening of $\Gamma_L=10\Gamma_{LR}$ and $\Gamma_R=0.1\Gamma_{LR}$ respectively, and the energy level of the two sites is $\epsilon=-\Gamma_{LR}$. The solid line denotes the Fermi function at the corresponding temperatures. We choose three different temperatures. In all the situations the function $n_d(\epsilon_d)$ fits well with the Fermi function. }
\end{figure}
Now let us study the local temperature of two sites coupled to an electron reservoir in thermal equilibrium with the different level broadening, in which we have shown that the zeroth law of thermodynamics is broken. 

The Hamiltonian describing the two sites, the reservoir and the auxiliary site is similar to Eq.~\ref{chainhamiltonian}, i.e.,
\begin{eqnarray}\nonumber\label{hamiltonian_measure}
\hat H &=& \sum_{k\sigma} \epsilon_k \hat c^\dag_{k\sigma} \hat c_{k\sigma} +  \sum_{\alpha' k\sigma} ( h_{\alpha'} \hat c^\dag_{k\sigma} \hat c_{\alpha'\sigma} + h.c.) \\ && \nonumber + \epsilon \sum_{\alpha'\sigma} \hat c^\dag_{\alpha'\sigma} \hat c_{\alpha'\sigma}+\epsilon_d \sum_\sigma \hat d^\dag_\sigma \hat d_\sigma \\ && + g' \sum_\sigma (\hat d^\dag_\sigma \hat c_{\alpha\sigma} +h.c.),
\end{eqnarray}
where $\hat d_\sigma$ denotes the annihilation operator at the auxiliary site, $\epsilon_d$ the energy level of the auxiliary site and $g'$ the coupling between the auxiliary site and the site $\alpha$ whose temperature is planned to be measured. For the given $\alpha$ ($\alpha=L,R$), the electron density at the auxiliary site as a function of $\epsilon_d$ can be calculated by the Keldysh Green's function method. We find that the spin-dependent electron density can be expressed as
\begin{equation}
 \begin{split}
 n_{d} =  \frac{1}{\pi } \int_{-\infty}^\infty d\omega  f & \left(|G^r_{d,L}|^2\Gamma_L+|G^r_{d,R}|^2\Gamma_R \right. \\ & \left.
+ 2\Gamma_{LR}\textbf{Re} (G^r_{d,L}G^{r*}_{d,R} ) \right),
\end{split}
\end{equation}
where $G^r_{d,L}$ and $G^r_{d,R}$ denote the retarded Green's functions between the auxiliary site and the site $L$ and $R$ respectively. These retarded Green's functions can then be obtained by solving the Dyson's equation. The electron density as a function of the energy level at the auxiliary site coupled to either the site $L$ or the site $R$ is shown in Fig.~5. We find that as $g'\to 0$ the data does fit well with the Fermi function at the temperature of the reservoir. This means that our method of defining the local temperature is appropriate in small quantum systems where the correlation between the system and its environment cannot be neglected.

It is well known that the partial density matrix of a small quantum system in thermal equilibrium is not a canonical one due to the coupling between the system and the environment. Our result shows that it is possible to assign a temperature to a system even its density matrix is not a canonical one.

\section{the definition of temperature in a system where the number of particles is conserved}

In above, we discuss the breakdown of the zeroth law in the systems consisting of identical fermions, in which we define the local temperature and the local chemical potential simultaneously because the systems are open. 

In this section, we turn to a closed system, in which the number of particles is conserved. Then we could temperarily forget the chemical potential and focus on the temperature. In details, the systems contain a series of spins with nearest-neighbor couplings. The reservoir is simulated by an infinite spin chain. A local magnetic field is supposed to split the energy levels of the spin up and spin down states in the systems. The Hamiltonian of the systems and the reservoir is expressed as
\begin{equation}
\begin{split}
 \hat H = & g \left( \sum_{i=-\infty}^{-\left(n+2\right)} + \sum_{i=\left(n+1\right)}^{\infty} \right) \left( \hat S^x_{i} \hat S^x_{i+1} + \hat S^y_{i} \hat S^y_{i+1} \right) \\ & + h_L \sum_{\alpha=x,y} \hat S^\alpha_{-\left(n+1\right)} \hat S^\alpha_{-n} +  h_R \sum_{\alpha=x,y} \hat S^\alpha_{n} \hat S^\alpha_{n+1} \\ & + g_L \sum_{i=-n}^{-2} \sum_{\alpha=x,y} \hat S^\alpha_{i} \hat S^\alpha_{i+1}  + g_R \sum_{i=1}^{n-1} \sum_{\alpha=x,y} \hat S^\alpha_{i} \hat S^\alpha_{i+1} \\ & + d \sum_{\alpha=x,y} \hat S^\alpha_{-1} \hat S^\alpha_{1}+ \epsilon \left( \sum_{i=-n}^{-1} + \sum_{i=1}^n \right) \hat S^z_i,
\end{split}
\end{equation}
where the sites $i$ with $|i| > n$ are in the thermal reservoir. The site at the infinity is supposed to be coupled to the site at the minus infinity, so that the reservoir is as a whole. The sites $i$ with $-n\leq i\leq -1$ are in the system $L$, and that with $1\leq i\leq n$ are in the system $R$. The $h_L$ and $h_R$ denote the coupling between the reservoir and the system $L$ and $R$ respectively, and $d$ denotes the coupling between the two systems. And $\epsilon$ denotes the Zeeman splitting energy in the systems.

The Hamiltonian can be re-expressed in terms of spinless fermions $\hat c_i^{(\dag)}$ through the Jordan-Wigner transformation:
\begin{equation}\label{jordan_spinhamiltonian}
\begin{split}
 \hat H = & \frac{g}{2} \left( \sum_{i=-\infty}^{-\left(n+2\right)} + \sum_{i=\left(n+1\right)}^{\infty} \right) \left( \hat c_i^\dag \hat c_{i+1} + h.c. \right) \\ & + \frac{h_L}{2} \left( \hat c^\dag_{-\left(n+1\right)} \hat c_{-n} + h.c.\right) +  \frac{h_R}{2} \left( \hat c^\dag_{n} \hat c_{n+1} + h.c.\right)  \\ & + \frac{g_L}{2} \sum_{i=-n}^{-2} \left( \hat c_i^\dag \hat c_{i+1} + h.c. \right)  + \frac{g_R}{2} \sum_{i=1}^{n-1} \left( \hat c_i^\dag \hat c_{i+1} + h.c. \right) \\ & + \frac{d}{2} \left( \hat c_{-1}^\dag \hat c_{1} + h.c. \right) + \epsilon \left( \sum_{i=-n}^{-1} + \sum_{i=1}^n \right) \hat c^\dag_i \hat c_i .
\end{split}
\end{equation}
We immediately notice that this model is exactly equivalent to the model described in Eq.~\ref{chainhamiltonian}, except that in Eq.~\ref{jordan_spinhamiltonian} the Hamiltonian of the reservoir is expressed in the real space while it is expressed in the momentum space in Eq.~\ref{chainhamiltonian}. And the spin degree of freedom is neglected in Eq.~\ref{jordan_spinhamiltonian}, since it is irrelevant as the electron-electron interaction is absent. The two end sites of the reservoir are at the positions of $\pm \left( n+1\right) $. And the level broadening at the edge sites of the system $L$ and $R$ are $\Gamma_L= \frac{h_L^2}{2g}$ and $\Gamma_R= \frac{h_R^2}{2g}$ respectively, due to the couplings to the reservoir. The local observable now becomes the spin density, which is directly related to the particle density by $\langle \hat S_i^z \rangle = \langle \hat c^\dag_i \hat c_i\rangle - \frac{1}{2}$.

Since this model is equivalent to the one that we discussed above, we get that the zeroth law is broken in this model at low temperatures. And the traditional way of defining the temperature fails for these closed systems. According to the above idea, the local temperature is defined by attaching an auxiliary spin at the position where we plan to measure the temperature. The auxiliary spin is a two-level system. A local magnetic field is applied to it so that its energy levels are splitted. Then the temperature of the auxiliary spin can be readout from the spin density as a function of the Zeeman splitting energy $\Delta$. The Hamiltonian describing the auxiliary spin and the coupling between it and the system is expressed as
\begin{equation}
 \hat H_a = \Delta \hat S^z_0 + g' \sum_{\alpha=x,y} \hat S^\alpha_0 \hat S^\alpha_1. 
\end{equation}
Without losing generality, we suppose that the auxiliary spin locates at the site $0$ and is coupled to the site $1$ in the system $R$. After the Jordan-Wigner transformation, the model changes into the one described in Eq.~\ref{hamiltonian_measure} except an unimportant interaction term. As $g'\to 0$, the spin density at the auxiliary site will be
\begin{equation}
 S^z_0(\Delta) = \frac{1}{2} \tanh \left( \frac{-\Delta}{2k_B T}\right).
\end{equation}
Obviously, the local temperature can be readout by plotting $S^z_0$ as a function of $\Delta$. 

\section{conclusions}

In this paper, we show the breakdown of the zeroth law of thermodynamics in small quantum systems at low temperatures. The traditional method of defining the temperature, in which the temperature is a quantity assigned to a set of systems in equilibrium with each other, fails in small quantum systems. Then we show a different way of defining the temperature. In this way we can obtain the local temperature of a system in equilibrium by attaching an auxiliary site to it. The temperature defined in this way is not an intrinsic quantity of the system's density matrix any more.

\end{document}